\documentclass[12pt]{iopart}
\usepackage{amssymb,graphicx,color}

 \newcommand{\kB}{k_{_B}}
 \newcommand{\Rg}{R_{_G}}
 \newcommand{\Dc}{D_{_c}}
 \newcommand{\ue}{\mathrm{e}}
 \newcommand{\ud}{\mathrm{d}}

 \newcommand{\RR}{\mathbf{R}}
 \newcommand{\rr}{\mathbf{r}}
 \newcommand{\qq}{\mathbf{q}}
 \newcommand{\ff}{\mathbf{f}}
 
 \newcommand{\bll}{\mathbf{\lambda}}
 \newcommand{\bxi}{\mathbf{\xi}}

 \newcommand{\one}{\mathbf{1}}

 \newcommand{\tU}{\tilde{U}}

 \newcommand{\tff}{\tilde{\mathbf{f}}}
 \newcommand{\dRR}{\dot{\mathbf{R}}}
 \newcommand{\ddRR}{\ddot{\mathbf{R}}}
 \newcommand{\drr}{\dot{\mathbf{r}}}

 \newcommand{\hmatL}{\hat{\Lambda}}
 
 \newcommand{\Rev}[1]{{\color{black} #1}}
 \newcommand{\rev}[1]{{\color{black} #1}}

\begin{document}

\title[Dynamic coarse-graining of polymer systems using mobility functions]
{Dynamic coarse-graining of polymer systems using mobility functions}

\author{B Li$^1$, K Daoulas$^2$, F Schmid$^1$}
\address{$^1$ Institut f\"ur Physik , Johannes
    Gutenberg-Universit\"at Mainz, 55099 Mainz, Germany}
\address{$^2$ Max-Planck Institut f\"ur Polymerforschung, 
    Ackermannweg 10, 55128 Mainz, Germany}

\begin{abstract}
We propose a dynamic coarse-graining (CG) scheme for mapping
heterogeneous polymer fluids onto extremely CG models in a dynamically
consistent manner. The idea is to use as target function for the
mapping a wave-vector dependent mobility function derived from the
single-chain dynamic structure factor, which is calculated in the
microscopic reference system. In previous work, we have shown that
dynamic density functional calculations based on this mobility
function can accurately reproduce the order/disorder kinetics in
polymer melts, thus it is a suitable starting point for dynamic
mapping. To enable the mapping over a range of relevant wave vectors,
we propose to modify the CG dynamics by introducing internal friction
parameters that slow down the CG monomer dynamics on local scales,
without affecting the static equilibrium structure of the system. We
illustrate and discuss the method using the example of infinitely long
linear Rouse polymers mapped onto ultrashort CG chains.  We show that our
method can be used to construct dynamically consistent CG models for
homopolymers with CG chain length $N=4$, whereas for copolymers,
longer CG chain lengths are necessary.
\end{abstract}

\noindent{\it Keywords\/}: polymer simulations; coarse-graining;
dynamics; friction; dynamic structure factor; dynamic density 
functional theory; mobility

\submitto{\JPCM}

\maketitle


\sloppy

\section{Introduction}
\label{Sec:Intro}

Mixing polymers of different types is a simple and inexpensive way to
create novel materials\cite{book_blends,book_multiphase}. However,
chemically different polymers usually do not mix well. Polymeric
composite materials  therefore tend to be heterogeneous on local 
scales and filled with internal interfaces, which largely determine
the resulting material properties\cite{book_interfaces}. The 
morphology of the materials depend on the history, i.e., the way 
they have been processed.  Understanding the dynamics of polymer
kinetics in inhomogeneous materials is thus crucial if one wants 
to understand and predict the structure and properties of the 
resulting materials.

Computer simulations are a powerful tool to study soft matter systems.
Due to the large size of the polymers and the even larger typical
length scales of the inhomogeneities, simulations in full atomistic
details are usually not possible, and using coarse-grained (CG) models
instead has a long and successful
history\cite{McCrackin_67}. 
In CG polymer models, monomers or groups of monomers are lumped into
one ''bead'' of simpler structure. Generic models offer insight into
universal features, and specific models with parameters adjusted to
concrete molecules are used for quantitative studies. Designing such
specific CG models requires the development of mapping procedures that
allow to derive the parameters of the CG models from the microscopic
static and dynamic features of the target systems \cite{baschnagel_00,mplathe_02,peter_09,peter_10,brini_13, noid_13,wagner_16}.

With respect to the static properties of equilibrium systems, such methods are by now well-established. Various protocols have been
proposed to derive effective potentials of coarse-grained models from
microscopic simulations by analyzing local correlations or force
distributions\cite{brini_13,noid_13}. In addition, established mesoscopic
concepts such as the Flory Huggins $\chi$-parameter\cite{book_multiphase},
the statistical segment length\cite{Doi}, or the Maier-Saupe
parameter\cite{MaierSaupe1959,MaierSaupe1960,deGennesBOOKLC} are used to map microscopic models (or experimental data) on continuum models and 
then back to extremely CG particle-based polymer
models\cite{Olsen2008,GrecoJCP2016,MappingMS2018}. In the latter case,
the target quantity in the CG parameter optimization is often the
static structure factor, and polymer theories like the random phase
approximation (RPA) or the self-consistent field theory (SCF) 
help to establish the connection between fine-grained and CG
models\cite{Morse2014,hannon_18,Matsen2019}.

Motivated by these successes, similar efforts are made to design
mapping and CG methods for polymer dynamics. In the earliest and still
very popular approach\cite{Tschop1998a,Tschop1998b}, the CG model is
simulated by standard molecular dynamics and a single time scale -- e.g.,
the time scale of diffusion -- is used to mapped the CG system onto the
fine-grained system. However, it has been long known through the work
of Mori and Zwanzig\cite{Zwanzig1961,Mori1965,Zwanzig2001}, that
coarse-graining has a much more
fundamental effect on the structure of the equations of motion:
Integrating out degrees of freedom invariably turns a Hamiltonian
system into a dissipative system with memory. Based on this insight,
several recent efforts have been devoted to deriving generalized
Langevin (GLE) models for polymer melts and solutions, using as target
quantities for the mapping the (absolute or relative) velocity
autocorrelation function of the center of mass\cite{li2015incorporation,li2016comparative,li2017computing,
wang2019implicit,wang2020data} of the
molecules. In most of these systems, whole polymer molecules
(typically relatively small star polymers) were mapped onto single CG
particles. Extending these concepts to CG models that map polymers on
CG chains with multiple sites is far from
trivial\cite{Chen2014,Ma2016,Lee2019}.  One approach that has been rather
successful in the case of  oligomer molecules was to integrate over
pair memory kernels and thereby derive dissipative particle
dynamics (DPD) friction constants for monomers\cite{Deichmann2018} 
\Rev{-- similar to earlier work by Hijon et al who used the Mori-Zwanzig 
formalism to construct DPD equations for CG particles representing whole 
star polymers\cite{Hijon2010}}.
However, it is not clear whether this approach will also work for
large molecules, where internal chain motion is a significant source
of memory and friction. An alternative route that is closer to the
static coarse-graining strategies developed for mesoscopic scales
would be to use the dynamic structure factor as a starting point for
mapping.  Such a strategy will be explored in the present paper.

We target systems containing polymers of large molecular weight, i.e.,
made of thousands of monomers, and CG strategies that map these
molecules onto much shorter chains of soft blobs. The dynamics of such
polymers is theoretically described as an overdamped motion in a
background medium created by the other polymers, e.g., Rouse, Zimm, or
reptation dynamics\cite{Doi}. Successful static CG strategies are
based on ''theoretically informed'' soft potentials that are derived
from static density functionals\cite{Laradji1994,Soga1995,Pagonabarraga,Daoulas2006,dePablo2009,Milano,MuellerReview,daoulas2012,Sevink,GrecoJCP2016} and reproduce key
quantities such as the $\chi$ parameter. Here we take a similar
approach, but use as mesoscopic reference theory the overdamped
dynamic density functional theory (DDFT). The standard Ansatz of
such a DDFT equation for polymers has the form\cite{Kawasaki1987,
Kawasaki1988,Fraaije1993,Fraaije1997,Kawakatsu1999,Muller2005,Wittkowski_2020}
\begin{equation}
 \label{eq:DDFT}
 \frac{\partial \rho_{\alpha}(\rr,t)}{\partial t} 
    = \sum_{\beta} \nabla_{\rr}\left[\int  d\rr^{\prime}
    \Lambda_{\alpha\beta}(\rr,\rr^{\prime})
    \nabla_{\rr^{\prime}} \mu_{\beta}(\rr^{\prime},t)\right].
\end{equation}
Here $\rho_\alpha(\rr,t)$ is the local density of component $\alpha$
at the position $\rr$, the quantity
$\nabla_{\rr^{\prime}}\mu_{\beta}(\rr^{\prime},t)$ is the
thermodynamic force acting on component $\beta$ at position $\rr'$,
and $\Lambda_{\alpha\beta}(\rr,\rr^{\prime})$ a non-local mobility
function that accounts, e.g., for chain connectivity effects.
Obviously, Eq.\ (\ref{eq:DDFT}) is Markovian and does not include
memory effects.  More general versions of (\ref{eq:DDFT}) that
includes a memory kernel $K(\rr,\rr^{\prime},t)$ have been proposed by
Semenov\cite{Semenov1986} and more recently by M\"uller and 
coworkers\cite{Wang2019polymers,Rottler2020}.  Eq.\ (\ref{eq:DDFT}) 
represents a Markovian approximation to the full  GLE which accounts 
for different relaxation times on different length  scales in an 
effective manner.  We will discuss this in more detail in the next section.

In previous work, we have devised a way to extract mobility functions
in polymer systems in a bottom-up fashion from fine-grained
simulations, using as input data the single chain dynamics structure
factor, $g(\qq,t) = \frac{1}{N} \langle \sum_{n,m=1}^N \ue^{i \qq
\cdot(\RR_n(t)-\RR_m(0))}\rangle$, where the sum $n,m$ runs over all
$N$ monomers of the chain, and $\RR_n(t)$ is the position of monomer
$n$ at time $t$.  Knowing $g(\qq,t)$, one can calculate the rescaled
single-chain mobility\cite{Mantha2020} in Fourier space as
\begin{equation}
\label{eq:Lambda}
\hmatL(\qq) = \frac{1}{\kB T N^2}\: g(\qq,0) G^{-1}(\qq)\: g(\qq,0)
\quad \mbox{with} \quad
G(\qq) = \frac{q^2}{N} \int_0^\infty \ud t \: g(\qq,t)
\end{equation}
In a homopolymer mixture containing polymers of type $\alpha$ (length
$N_\alpha$) in the number concentration $c_\alpha$, the total mobility
function is then given by\cite{Mantha2020} 
\begin{equation}
\label{eq:mobility_total}
\Lambda_{\alpha\beta} = \delta_{\alpha \beta} 
   \sum_{\alpha} c_\alpha N_\alpha^2 \hmatL^{(\alpha)}
\end{equation}
The generalization to block copolymers is straightforward
\cite{Mantha2020,Schmid2020}.  In our previous work, we have shown that a DDFT
(\ref{eq:DDFT}) based on this approach can accurately reproduce the
kinetic evolution of block copolymer melts after sudden changes of the
$\chi$-parameter, when compared to fine-grained reference simulations. 

These successes suggest that the mobility functions $\hmatL(\qq)$
should be a suitable target for dynamic CG schemes that map
fine-grained models to particle-based CG models.  In the present
paper, we will investigate this possibility. We will show that a naive
''mapping'' based on matching a single time scale fails to
reproduce the kinetics on both local and polymeric length scales.
This can partly be remedied by modifying the internal polymer dynamics
in the CG model. We will present a simple approach to do so and
discuss its limitations.

The remainder of the paper is organized as follows: In the next
section, we briefly discuss the background of the method. We first
introduce the mobility function in some more detail, and then discuss
finite chain length effects and the ensuing problems with simple time
mapping.  In Section \ref{Sec:ModifiedDynamics}, we propose a method
to modify the CG dynamics whithout affecting the static properties of
the systems and show results for an extremely coarse-grained polymer.
We close with a brief summary in Section \ref{Sec:Summary}.

\section{Background}
\label{Sec:Background}

\subsection{Single chain dynamic structure factor and mobility function}
\label{Sec:Mobility}

To set the frame, we begin with a brief derivation of Eq.\
(\ref{eq:Lambda}). It follows the spirit of the derivation presented in
Ref.\ \cite{Mantha2020}, but specifically highlights the relation between the
mobility function and the corresponding single-chain memory kernel. For
simplicity, we again consider homopolymers.

We make two important assumptions. First, we assume that we can
determine the mobility function in a homogeneous \rev{(compressible)}
reference system (i.e., it is transferable to inhomogeneous systems),
and second we take a mean-field approach. We consider a tagged polymer
that moves in the average background potential provided by the other
chains of the reference system. Since the reference system is
homogeneous, we can write a generalized DDFT equation for the monomers
of the tagged polymer as follows:
\begin{equation}
 \label{eq:GLE_tagged}
 \frac{\partial \rho^{(s)}(\rr,t)}{\partial t} 
    = \nabla_{\rr} 
        \int  \ud \rr^{\prime} \:
        \int_{-\infty}^t \ud s \: 
        K^{(s)}(\rr-\rr^{\prime},t-s) \: 
        \nabla_{\rr^{\prime}} \mu^{(s)}(\rr^{\prime},s),
\end{equation} 
which in Fourier space reads
\begin{equation}
 \label{eq:GLE_tagged_qq}
 \frac{\partial \rho^{(s)}(\qq,t)}{\partial t} 
    = - q^2 \int_{-\infty}^t \ud s \:
 K^{(s)}(\qq,t-s) \: \mu^{(s)}(\qq,s),
\end{equation}
(using the convention
$f(\qq) = \int \ud \rr \: \ue^{i \qq \cdot \rr} f(\rr)$
for the Fourier transform), where
$\mu^{(s)}(\qq) = V \delta F^{(s)}/\delta \rho^{(s)}(-\qq)$
is derived from the free energy $F^{(s)}$ of the single
tagged chain system. 
Eqs.\ (\ref{eq:GLE_tagged}) account for memory effects {\em via}
the single-chain memory kernel $K^{(s)}(\tau)$. They do not
include corresponding correlated stochastic currents, 
but these could be added easily and would drop out in the next step 
of the derivation.

The single chain structure factor is then given by
$g(\qq,t) = \frac{1}{N} \langle \rho^{(s)}(\qq,t) \rho^{(s)}(-\qq,0)
\rangle$, where $\langle \cdot \rangle$ denotes the thermal
average over chain configurations. This results in the following
equation for $g(\qq,t)$: 
\begin{equation}
 \label{eq:GLE_g_1}
 \frac{\partial g(\qq,t)}{\partial t} 
    = - \frac{q^2}{N}\int_{-\infty}^t \ud s \:
          K^{(s)}(\qq,t-s) \: 
         \langle \mu^{(s)}(\qq,s) \rho^{(s)}(- \qq,0) \rangle.
\end{equation}
To calculate $\mu^{(s)}$,  we linearize the tagged chain free energy 
$F^{(s)}$ and expand it in powers of the tagged monomer density
$\rho^{(s)}$, \footnote{In Ref.\ \cite{Mantha2020}, \Rev{the corresponding
equation, Eq. (16), contains} an additional erroneous factor $1/V$.} 
\begin{equation}
F^{(s)} = \mbox{const.} + 
\frac{\kB T}{2 N} \sum_{\qq} \rho^{(s)}(-\qq) \: g^{-1}(\qq,0) \:
\rho^{(s)}(\qq) + \cdots
\end{equation}
\Rev{By truncating this equation at the second order, we implicitly
assume that the chain conformations stay close to equilibrium and
are not strongly distorted. Taking the derivative}
 \footnote{In Ref.\ \cite{Mantha2020} (before Eq. (17), the factor $V$ is
 missing.},  
$\mu^{(s)}(\qq,t) = \frac{\kB T V}{N} g^{-1}(\qq,0)\rho^{(s)}(\qq,t)$,
\Rev{and inserting it} in Eq. (\ref{eq:GLE_g_1}), we obtain
\begin{equation}
 \label{eq:GLE_g_2}
 \frac{\partial g(\qq,t)}{\partial t} 
    = - \frac{q^2 \kB T V}{N}\int_{-\infty}^t \ud s \:
          K^{(s)}(\qq,t-s) \: 
          g^{-1}(\qq,0) \: g(\qq,s).
\end{equation}
Next we carry out a one-sided Fourier transform in the time domain
\begin{equation}
 \label{eq:GLE_g_3}
  i \omega \tilde{g}(\qq,\omega) - g(\qq,0)
  = - \frac{q^2 \kB T V}{N} \tilde{K}^{(s)}(\qq, \omega)
    \: \tilde{g}^{-1}(\qq,0) \: \tilde{g}(\qq,\omega),
\end{equation}
which finally allows to calculate $\tilde{K}^{(s)}(\qq,\omega)$ as
\begin{equation}
\label{eq:Ks}
\tilde{K}^{(s)}(\qq,\omega)
 = \frac{N}{\kB T V q^2}
  \big( g(\qq,0) - i \omega \tilde{g}(\qq,\omega) \big) \:
  \tilde{g}^{-1}(\qq,\omega) \: g(\qq,0).
\end{equation}

\rev{These equations can easily be generalized to copolymers
containing different types of monomers replacing $\rho$ and $\mu$ with
vectors, and $g$, $K$ with matrices.} We emphasize that
$K^{(s)}(\qq,\tau)$ represents a {\em single-chain} memory kernel,
which describes the self-diffusion of the tagged chain.  Wang et
al\cite{Wang2019polymers} have recently calculated the {\em
collective} memory kernel for incompressible block copolymer melts
within the random phase approximation \rev{and, interestingly, obtained
essentially the same expression (with a modification due to the
incompressibility condition). The exact collective memory kernel can be
obtained from simulations using a similar expression than
(\ref{eq:Ks}), with the single chain structure factor replaced by the
collective structure factor.} For the purpose of dynamical mapping, it
is more convenient to use the single chain structure factor as target
quantity, since it can be accessed more easily over the whole range of
$\qq$ vectors even from fine-grained simulations of very small
systems. \Rev{A second advantage is that the single-chain structure
factor is much less affected by dynamic slowdown close to phase
transitions, which may occur due to slow collective critical or
near-critical fluctuations \cite{Morse2019}.  This makes it easier to
justify the Markovian approximation described below.}

In order to derive Eq.\ (\ref{eq:DDFT}) with (\ref{eq:Lambda}) from
Eq.\ (\ref{eq:GLE_g_2}) with (\ref{eq:Ks}), we apply a Markovian
approximation\cite{Hijon2010,Deichmann2018} and replace
$K^{(s)}(\qq,\tau)$ by $K^{(s)}(\qq,\tau) \approx \Lambda^{(s)}(\qq)
\: 2 \delta(\tau)$, where the single-chain mobility is the integral
over the memory kernel
\begin{equation}
\label{eq:lambdas}
\Lambda^{(s)}(\qq) = \int_0^\infty \ud \tau \: K^{(s)}(\qq,\tau)
  = \tilde{K}^{(s)}(\qq,0)
\end{equation}

Inserting Eq.\ (\ref{eq:Ks}), identifying $G(\qq) = \frac{q^2}{N}
\tilde{g}(\qq, 0)$ and rescaling\footnote{\Rev{In the corresponding expressions 
in Ref.\ \cite{Mantha2020} (after (14) and before (19)) a factor $V$ is
missing.}}
via  $\hmatL = \Lambda^{(s)} \frac{V}{N^2}$,
we recover Eq.\ (\ref{eq:Lambda}). Within the Markovian approximation,
$g(\qq,t)$ decays exponentially (see Eq.\ (\ref{eq:GLE_g_1})): The
multiple relaxation times contributing to the memory kernel are
replaced by one effective relaxation time, which is, however, a
function of  $\qq$.  {\em Via} this $\qq$-dependence, one still
accounts, to some extent, for the spectrum of characteristic
relaxation modes in polymers. As we have seen in our previous
work\cite{Mantha2020}, this seems to be sufficient to reproduce the
ordering/disordering kinetics in melts at a quantitative level.

The mobility function $\hmatL(\qq)$ can thus be used to characterize
the polymer dynamics in a fine-grained system. Based on this insight,
we propose to use it as target function for a dynamically consistent
mapping of \Rev{fine-grained} systems onto \Rev{CG} systems. As we
shall see in the next subsection, such a mapping is far from trivial.

\subsection{Chain length effects on the mobility function}
\label{Sec:Mobility_finite size}

In a previous publication\cite{Schmid2020}, we have derived an expression for
the single-chain mobility function of ideal infinitely long chains in
the Rouse regime. The result was lengthy and shall not be repeated
here.  However, simple expressions were obtained for the limiting cases
of very small or very large length scales. For homopolymers, we get
\begin{eqnarray}
 \label{eq:mobility_qlarge}
q \Rg \to \infty &:&
\hmatL(\qq) \to \frac{\Dc}{\kB T} \cdot 0.279 \\
q \Rg \to 0 &:&
\hmatL(\qq) \approx \frac{\Dc}{\kB T} \cdot 
 \big( 1- \frac{(q \Rg)^2}{3} \big),
 \label{eq:mobility_qsmall}
\end{eqnarray}
where $\Rg$ is the radius of gyration, and $\Dc$ the diffusion
constant of the chain.

\begin{figure}[tbp]
    \centering
    \includegraphics[width=0.95\textwidth]{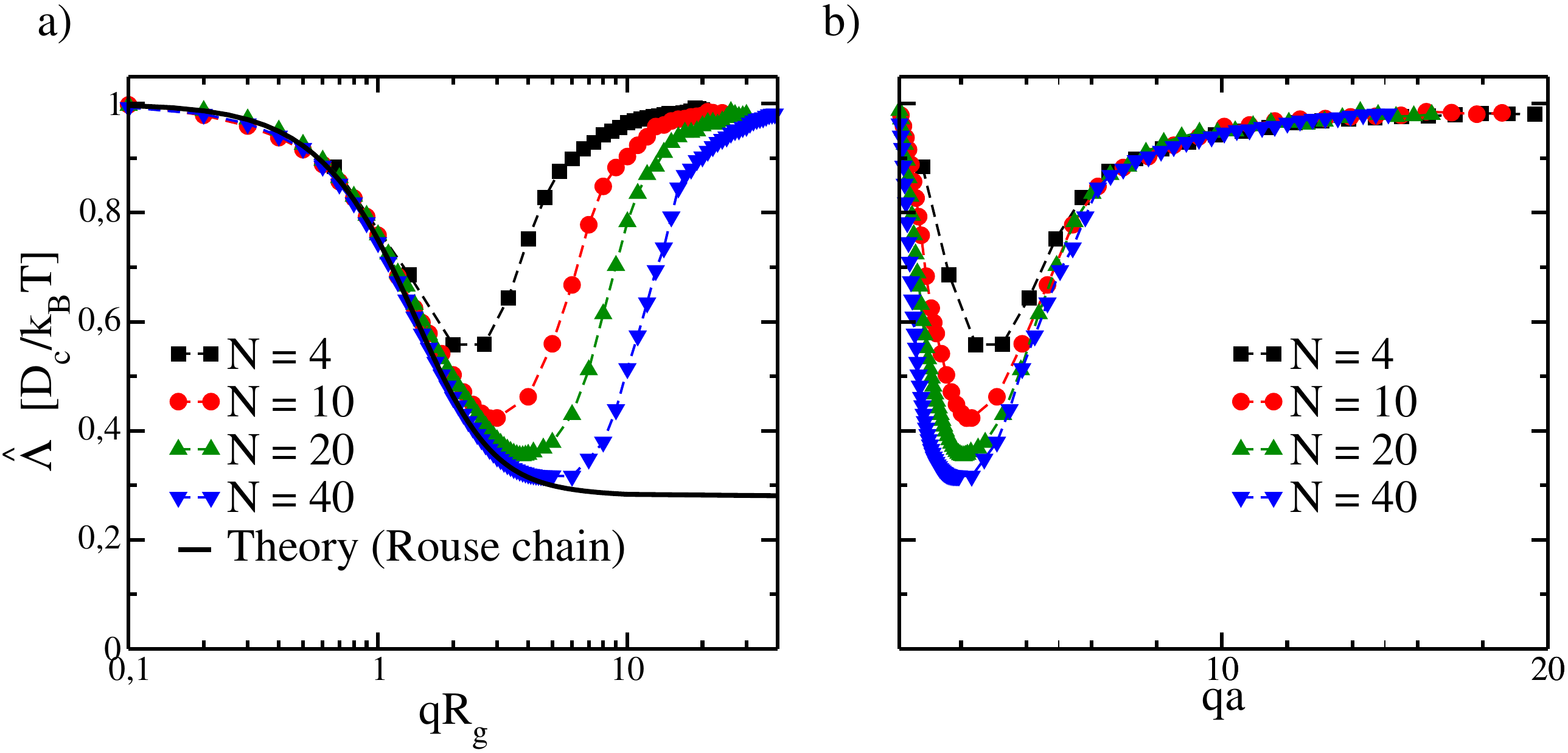}
    \caption{Rescaled mobility functions $\Lambda$ for polymer chains
    with different chain length as indicated. a) Logarithmic plot
    versus $q \Rg$. b) Linear plot versus $q a$, where $a$ is the
    statistical segment length. The black solid line shows
    the theoretical results from Ref.\ \protect\cite{Schmid2020} }
    \label{fig:fig_obd}
\end{figure}

In CG polymer models, one represents polymers by relatively short,
possibly very short chains. This turns out to have a significant
impact on the mobility function. To investigate the chain length
effects, we have carried computer simulations of spring-bead chains
with harmonic bond potentials and different numbers of beads. Apart
from being connected by bonds, monomers do not interact with each
other. They move according to overdamped Brownian dynamics equations
with a monomer friction constant $\zeta$. To determine the mobility
functions from the simulation data, we first determine the single chain
structure factor $g(\qq,t)$ from the simulation trajectories and
then evaluate the integral $G(\qq)$ and finally $\hmatL$ according to Eq.\ (\ref{eq:Lambda}), applying an extrapolation procedure as described
in Ref.\ \cite{Mantha2020} if necessary. The results for different
chain lengths are presented in Fig.\ \ref{fig:fig_obd}. To normalize
the data, the mobilities are divided by the respective polymer
diffusion constants $D_c = 1/\kB T \zeta N$.  In Fig.\
\ref{fig:fig_obd} a), we also show the theoretical result for
infinitely long Rouse polymers\cite{Schmid2020}.

The simulation data agree well with the theory for small $q$.  At
larger $q$, however, they deviate. Different from Rouse polymers, the
mobility functions of finite chains are nonmonotonic. They start from
$\hmatL(0)=\Dc/\kB T$ and first decay, initially closely following the
theoretical curve, but then assume a minimum and grow again, until
they reach the original value, $\hmatL(q) = \Dc/\kB T$ at $q \to
\infty$.  In the small $q$ regime, the curves for different chain
lengths collapse onto each other if plotted against $q \Rg$; in the
large $q$ regime, they collapse if plotted as a function of $q$ only
(made dimensionless by multiplying with the the statistical segment
length $a$).

In the DDFT (Eq.\ (\ref{eq:DDFT})), the asymptotic large $q$ behavior of
$\hmatL$ describes that expected for a fluid of monomers which
move independently with the diffusion constant $D_0 = \hmatL(\infty) N
= 1/\kB T \zeta$\cite{Muller2005,Qi2017}. Hence, we observe a crossover from a
collective ''chain mobility'' to a ''monomer mobility'' in chains with
finite length $N$. The crossover point (the position of the minimum)
scales roughly like $(q\Rg)_c \sim N^{1/3}$ as a function of
chain length.  This seems to suggest that the crossover wavelength is
determined by the average distance $d$ of monomers in the coil, which
is set by the local density, $d \sim \rho^{1/3}$ with $\rho =
N/\Rg^3$. In the limit of infinite chain length, the crossover
point $(q \Rg)_c$ moves to infinity. However, the value of
the bare wavevector at the crossover, $(qa)_c$, moves to zero
for infinite chain length.

The reason why the mobility of the finite chain at large $(q \Rg)$
differs from that of the infinite chain can be rationalized as
follows: In the regime $1 \ll (q \Rg) \ll (q \Rg)_c$, the local
mobility is dominated by the collective motion of whole chain portions
with a locally scale invariant conformations. The effective friction
of such a ''wad'' is reduced, compared to that of a monomer, and can
be calculated from its local self-similar structure\cite{Schmid2020}.  
On the other hand, on ultra-short length scales with $(q \Rg)_c \ll (q \Rg)$,
the effect of chain connectivity becomes negligible and monomers
diffuse individually. The two regimes (''wad'' diffusion and monomer
diffusion) are well separated in real polymer systems.  However, in CG
model systems of short chains, they move closer to each other and
overlap.

When devising dynamic mapping schemes for such extremely CG polymers
systems that cover kinetic processes, one is thus faced with a
fundamental problem: It is impossible to accurately represent dynamic
processes on both global and local (''wad'') length scales with simple
time scale matching.  If one uses the time scale of chain diffusion
for time mapping, the time scales of local ordering, e.g., at
interfaces, are overestimated by a factor of roughly 3.6. On the other
hand, if maps the time scale of local ordering, the global chain
diffusion is underestimated.

We should note that related finite chain length effects are also observed 
in the static structure factor, $g(\qq,0)$, although they are much
less dramatic: For infinitely long chains, $1/N g(\qq,0)$ drops from
1 (at $\qq = 0$) to zero at large $\qq\Rg \to \infty$, whereas it 
levels off at $1/N$ for finite chains. In principle, this can be
corrected by an appropriate backmapping procedure~\cite{Zhang2014AML}, 
i.e., restoring  structure in the coarse-grained beads in retrospect. 
In the case of the dynamics, a different approach must be taken.

\section{Adapting the CG polymer dynamics on multiple length scales}
\label{Sec:ModifiedDynamics}

We will now propose a way to adjust the CG dynamics of in a CG polymer
system such that it has the same mobility function than the target
system of large polymers over the whole range of $\qq$ vectors up to
$(q \Rg)_c$. The idea is slow down the internal modes, such that the
CG monomers effectively have the mobility of a ''wad'', without
changing the diffusion constant of the whole chains and the static
structure of the chains. To this end, we have to introduce different
friction constants for internal modes and global diffusion.

\subsection{Method}
\label{Sec:Methods}

We consider linear Gaussian chains of length $N$ with global chain
friction $\gamma_t$. Our goal is to devise a modified dynamical model
that allows for different internal friction constants while not
affecting the static behavior of the chain. The diffusion constant of
the whole chain will be kept fixed.

The monomer coordinates are given by $\RR_i(t)$, and the total
potential is given by $U[\{ \RR_i\}]$. Thus the force acting on
monomer $i$ is given by $\ff_i = - \nabla_{\RR_i} U$.  The center of
mass of the chain is given by $\RR_t(t) = \frac{1}{N} \sum_i \RR_i$
and the total force acting on all monomers is $\ff_t(t) = \sum_i
\ff_i(t)$.

\subsubsection{Overdamped Brownian dynamics with two friction
constants.}

The simplest Ansatz is to introduce two friction constants, one for
the center of mass motion of the chain and one for the relative motion
with respect to the center of mass. We will illustrate this approach
using the example of a overdamped Brownian dynamics.  We introduce
alternative coordinates $\RR_t$ (center of mass) and $\rr_i = \RR_i -
\RR_t$ (internal coordinates), i.e., $\sum_i \rr_i = 0$. Rewriting the
potential energy as a function of these coordinates, we obtain a new
potential function
\begin{equation}
\label{eq:tU}
\tU[\RR_t,\{\rr_i\}] = U[\{\RR_t + \rr_i\}]. 
\end{equation}
To reproduce the identical static averages, the generalized forces $\tff_i$ 
acting on coordinates $\RR_t$, $\rr_i$ are derived from $\tU$ with an 
additional Lagrange multiplier $\bll$ (a vector) 
that accounts for the constraint $\sum_i \rr_i \equiv 0$:
\begin{eqnarray}
\tff_t &=& - \nabla_{\RR_t} \tU 
 = \sum_i (- \nabla_{\RR_i} U) \frac{\partial \RR_i}{\partial \RR_t}
 = \sum_i \ff_i = \ff_t \\
\tff_i &=& - \nabla_{\rr_i} (\tU + \bll \cdot \sum_i \rr_i)
   = \ff_i - \bll
\end{eqnarray}
The constraint forces must be chosen such that the constraint is
fulfilled at all times.
The dynamical equations are overdamped Langevin equations
\begin{eqnarray}
\dRR_t &=& \gamma_t \tff_t + \bxi_t = \gamma_t \ff_t + \xi_t \\
\drr_i &=& \gamma_m \tff_i + \bxi_i = \gamma_m (\ff_i - \bll) + \bxi_i 
\label{eq:drr}
\end{eqnarray}
with inverse friction constants $\gamma_t$ and $\gamma_m$. The value
of $\gamma_t$ is chosen such that the chain has the desired diffusion
constant. The value of $\gamma_m$ can be used for mapping the dynamics
on short scales. The variables $\bxi_t, \bxi_i$ describe uncorrelated
Gaussian noise with mean zero ($\langle \xi_\alpha \rangle = 0$) which
satisfy the fluctuation-dissipation relation, i.e., $\langle \bxi_t(t)
\bxi_t(t') \rangle = 2 \kB T \gamma_t \one \delta(t-t')$, $\langle
\bxi_i(t) \bxi_i(t') \rangle = 2 \kB T \gamma_m \one \delta(t-t')$, and
$\langle \bxi_\alpha(t) \bxi_\beta(t') \rangle = 0 $ for $\alpha \neq
\beta$.  From the constraint $\sum_i \rr_i\equiv 0$, we derive $\sum_i
\drr_i \equiv 0$, which allows to express $\bll$ as $\bll =
\frac{1}{N}(\ff_t + \frac{1}{\gamma_m} \sum_i \bxi_i)$, hence
Eq.\ (\ref{eq:drr}) reads
\begin{equation}
\label{eq:drr2}
\drr_i = \gamma_m (\ff_i - \frac{1}{N} \ff_t) 
         + \bxi_i - \frac{1}{N} \sum_j \bxi_j.
\end{equation}
This finally yields the modified equations of motion for monomers $\RR_i$:
\begin{equation}
\dRR_i = \gamma_m \ff_i + \gamma_{t,\mbox{\tiny eff}} \ff_t + \eta_i
\quad \mbox{with} \quad
\gamma_{t,\mbox{\tiny eff}} = \gamma_t - \frac{1}{N} \gamma_m
\label{eq:dr_modified}
\end{equation}
where $\eta_i = \bxi_i  + (\bxi_t - \frac{1}{N}\sum_j \bxi_j)$.  Note
that $\eta_i$ is again a correlated Gaussian distribution noise with
correlation matrix $\langle \eta_i(t) \eta_j(t') \rangle = 2 \kB T
\delta(t-t') [\gamma_m \delta_{ij} + \gamma_{t,\mbox{\tiny eff}} ]$.
We recover the regular equations for linear Rouse polymers in the case
$\gamma_t = \frac{1}{N} \gamma_m$. 

\subsubsection{Generalizations.}
\label{Sec:Generalization}

The above modifications can also be applied to regular Langevin
dynamics (with inertia). For beads of mass $m$, we obtain the modified
equation of motion
\begin{equation}
m \ddRR_i = \ff_i - \zeta_m \dRR_i - \zeta_{t,\mbox{eff}} \dRR_t
   + \ff^{R}_i(t)
   \label{eq:drr_modified}
\end{equation}
where $\zeta_m = \gamma_m^{-1}$, $\zeta_{t,\mbox{eff}} = \frac{1}{N}
\gamma_t^{-1}- \gamma_m^{-1}$, and $\ff^R_i(t)$ is a Gaussian
distributed stochastic force with correlation matrix $\langle
\ff^R_i(t)\ff^R_j(t) \rangle = 2 \kB T \one \delta(t-t') [\zeta_m
\delta_{ij} + \frac{1}{N} \zeta_{t,\mbox{eff}}]$. As in Eq.\
(\ref{eq:drr2}), it can be implemented as a linear combination
of uncorrelated random forces $\ff^R_i = \theta_i +
\frac{1}{N}(\theta_t - \sum_j \theta_j)$ with $\langle \theta_t(t)
\theta_t(t') \rangle = 2 \kB T \gamma_t^{-1} \one \delta(t-t')$ and $
\langle \theta_i(t) \theta_j(t') \rangle = 2 \kB T \zeta_m \one
\delta(t-t')$.

Extensions to modified dynamical models with more than one internal
friction constants are straightforward. For future reference, we
briefly describe the resulting equations for a hierarchical model with
three friction constants. We separate the polymer into two blocks of
equal length, A and B,  such that the block A comprises monomers $\RR_i$
with $i \in \{1,...,N/2\} := S_A$ and the block B monomers $\RR_i$
with $i \in \{ N/2+1,...,N\} =: S_B$. We distinguish between the
forces $\ff_i$ acting on monomers $i$, the force $\ff_t = \sum_i
\ff_i$ acting on the whole chain, and the forces $\ff_{A,B} = \sum_{i
\in S_{A,B}} \ff_i$ acting on the individual blocks A and B. As
generalized coordinates, we choose the center of mass $\RR_t$ of the
full chain, the center of masses $\rr_{A,B}$ of the blocks relative to
$\RR_t$, and the coordinates $\rr_i$ of monomers relative to
$\rr_{A,B}$. The motion of $\RR_t$, $\rr_{A,B}$, and $\rr_i$ are
associated with separate inverse friction constants $\gamma_t,
\gamma_b$, and $\gamma_m$.  Following the same program as in the
previous subsection, we obtain the following dynamical equations for
monomers $i$ belonging to the block $\alpha(i)$ ($\alpha = A,B$)
(overdamped regime):
\begin{equation}
\dRR_i = \gamma_m \ff_i + \gamma_{b,\mbox{\tiny eff}} \ff_{\alpha(i)}
       + \gamma_{t,\mbox{\tiny eff}} \ff_t + \eta_i
\end{equation}
with $\gamma_{b,\mbox{\tiny eff}} = \gamma_b - \frac{2}{N} \gamma_m$,
$\gamma_{t,\mbox{\tiny eff}} = \gamma_t = \frac{1}{2} \gamma_b$, 
where the Gaussian noise is correlated according to
\begin{equation}
\langle \eta_i(t) \eta_j(t') \rangle = 2 \kB T
\delta(t-t') \big[ \gamma_m \delta_{ij} 
  + \gamma_{b,\mbox{\tiny eff}} \delta_{\alpha(i), \alpha(j)}
  + \gamma_{t,\mbox{\tiny eff}} 
\big].
\end{equation}
As in the previous examples, it can again be conveniently calculated
as a sum over uncorrelated Gaussian noise terms.  Similar equations
can be derived for other distributions of friction constants.

\subsection{Results}
\label{Sec:Results}

To evaluate our proposed approach, we consider an extreme test case
and attempt to map a Rouse polymer onto very short discrete Gaussian
chains (length $N$). As a preliminary remark, we note that the
asymptotic value of the mobility function $\hmatL(q)$ in the limit $q
\to \infty$ is bounded from below by the corresponding value for
chains with frozen conformations, where the chains only move as a
whole, i.e.\cite{Maurits1997,Qi2017}  $\hmatL(q)_{\mbox{\tiny frozen}} =
\frac{D_c}{\kB T} \frac{1}{N} g(q,0) \stackrel{q \to \infty}{\to}
\frac{D_c}{\kB T} N^{-1}$. In order to be able to implement the
limiting behavior of $\hmatL(q)$ for Rouse polymers given by Eq.\
(\ref{eq:mobility_qlarge}), CG chains must thus have a minimum length
of $N=4$. Hence we will evaluate CG systems of Gaussian tetramers.
We employ modified dynamics with two friction constants as
described in the previous section, Section~\ref{Sec:Methods}. 

\begin{figure}[btp]
    \centering
    \includegraphics[width=0.95\textwidth]{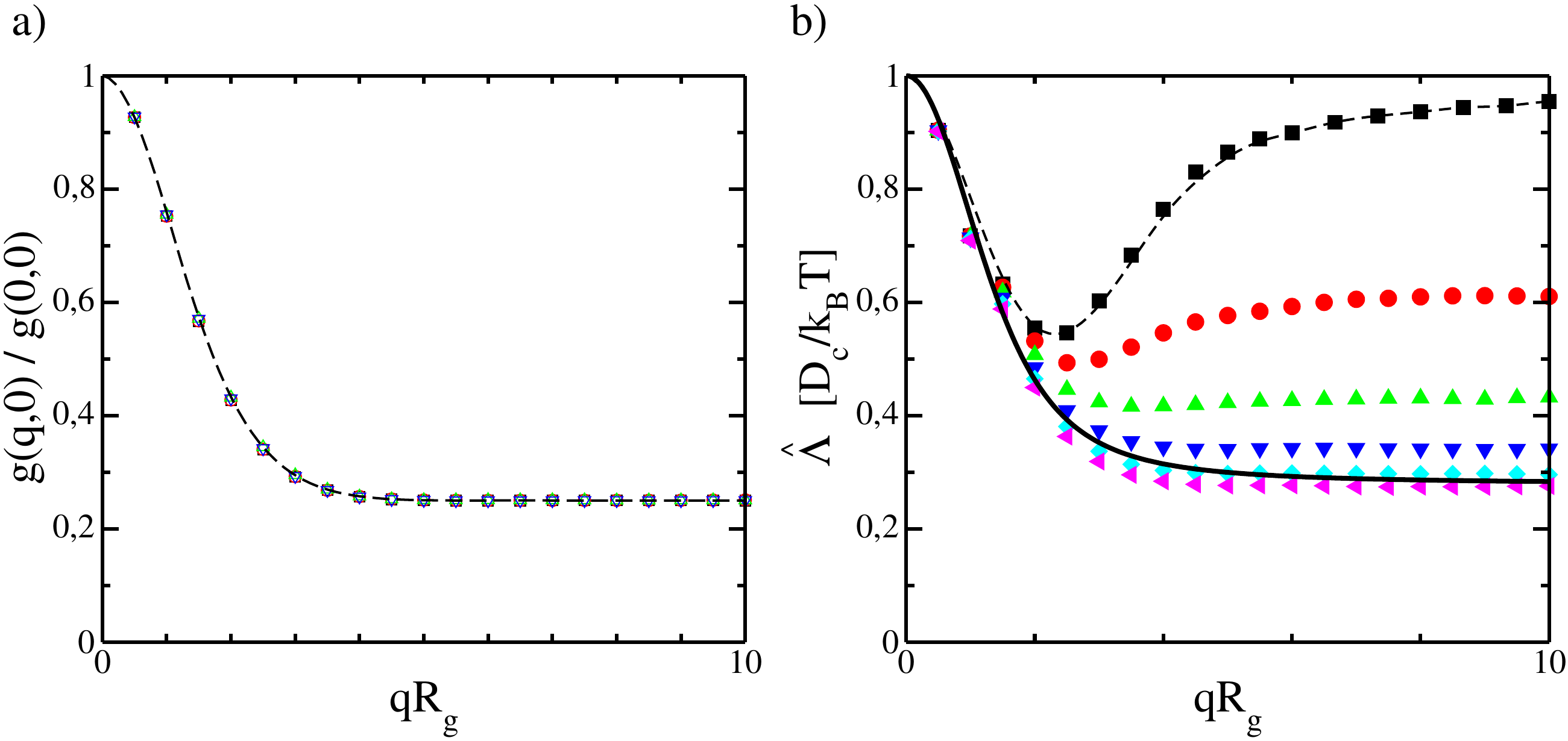}
    \caption{a) Normalized static structure factors $g(q,0)$ for
    chains with length $N=4$. Black dashed line shows the results from
    traditional overdamped Brownian dynamics simulations, the symbols
    those from modified dynamics approaches with two inverse friction
    constants $\gamma_m=N\gamma_t, N\gamma_t/16$ and three inverse
    friction constants with two sets : ($\gamma_b = N\gamma_t/2,
    \gamma_m = N\gamma_t $) and ($\gamma_b = N\gamma_t/16, \gamma_m =
    N\gamma_t/16$).  
    b) Rescaled mobility functions $\hmatL(q)$ for the same chains
    from modified dynamics with two friction constants.  For
    comparison the thin dashed line shows the results from traditional
    simulations, the thick solid line the theoretical results. For
    modified dynamics, the inverse relative monomer friction
    $\gamma_m$ is decreased from top to bottom: $\gamma_m =
    N\gamma_t$, $N\gamma_t/2$, $N\gamma_t/4$, $N\gamma_t/8$,
    $N\gamma_t/16$, $N\gamma_t/32$.   
    } \label{fig:fig_fullchain}
\end{figure} 

We first verify that the static behavior of the chain is not changed
by the modified dynamics model. We characterize the static properties
by the static structure factor, i.e., $g(q,t)$ at $t=0$.
Fig.~\ref{fig:fig_fullchain}a) shows the static structure factors
$g(q,0)$ for the polymer chains moving according to the modified
dynamics approaches with two friction constants ($\gamma_m=N\gamma_t,
N\gamma_t/16$) and three friction constants (two sets: $\gamma_b =
N\gamma_t/2, \gamma_m = N\gamma_t $ and $\gamma_b = N\gamma_t/16,
\gamma_m = N\gamma_t/16 $). The results from traditional overdamped
Brownian simulations are shown as black solid curves for comparison.
The agreement is excellent. Clearly the modified dynamics approaches
do not change the static behavior of the chain. 

Next we calculate the mobility function using the modified dynamics
approach with two friction constants. Fig.~\ref{fig:fig_fullchain}b)
shows the resulting mobility functions calculated for six values of
inverse relative monomer friction ($\gamma_m=N\gamma_t, N\gamma_t/2,
N\gamma_t/4, N\gamma_t/8, N\gamma_t/16$, and $ N\gamma_t/32$).  The
parameter $\gamma_t$, which sets the diffusion constant of the whole
chain, is kept fixed. Additionally shown is the result from
traditional simulations (black dashed line), and the theoretical
results for Rouse polymers from our previous paper, Ref.\ \cite{Schmid2020}
(black solid line).  

At small $q\Rg$ ($q\Rg<2$), the relative monomer friction has no
influence on the chain mobility functions. The data follow the
theoretical curve for Rouse polymers. The overall translational motion
of the chain dominates at small $q\Rg$. At intermediate and large
$q\Rg$, the internal relaxation becomes important. If one decreases
$\gamma_m$, the mobility function decreases. At
$\gamma_m \approx N\gamma_t/16$, the data for the CG chain match 
those of the Rouse polymer.  Hence, we can indeed obtain the target
mobility function in modified dynamics simulations by tuning
$\gamma_m$. This is the central message of the present paper. 

However, the approach also has limitations. This becomes apparent when
looking at the partial mobility functions for chain blocks, which is
important for dynamical studies of block copolymer ordering and
disordering\cite{Mantha2020,Schmid2020}. To illustrate this, we split our
ultrashort chain ($N=4$) in two symmetric blocks A and B of length $N=2$ 
(see Sec.\ \ref{Sec:Generalization}) and evaluate separately their mobility
functions $\hmatL_{AA}(q), \hmatL_{BB}(q)$ as well as the
cross-mobility $\hmatL_{AB}(q)$.  The same quantities can be
calculated semi-analytically for Rouse polymers using the expressions
given in our previous work, Ref.\ \cite{Schmid2020}. 

\begin{figure}[btp]
    \centering
    \includegraphics[width=0.95\textwidth]{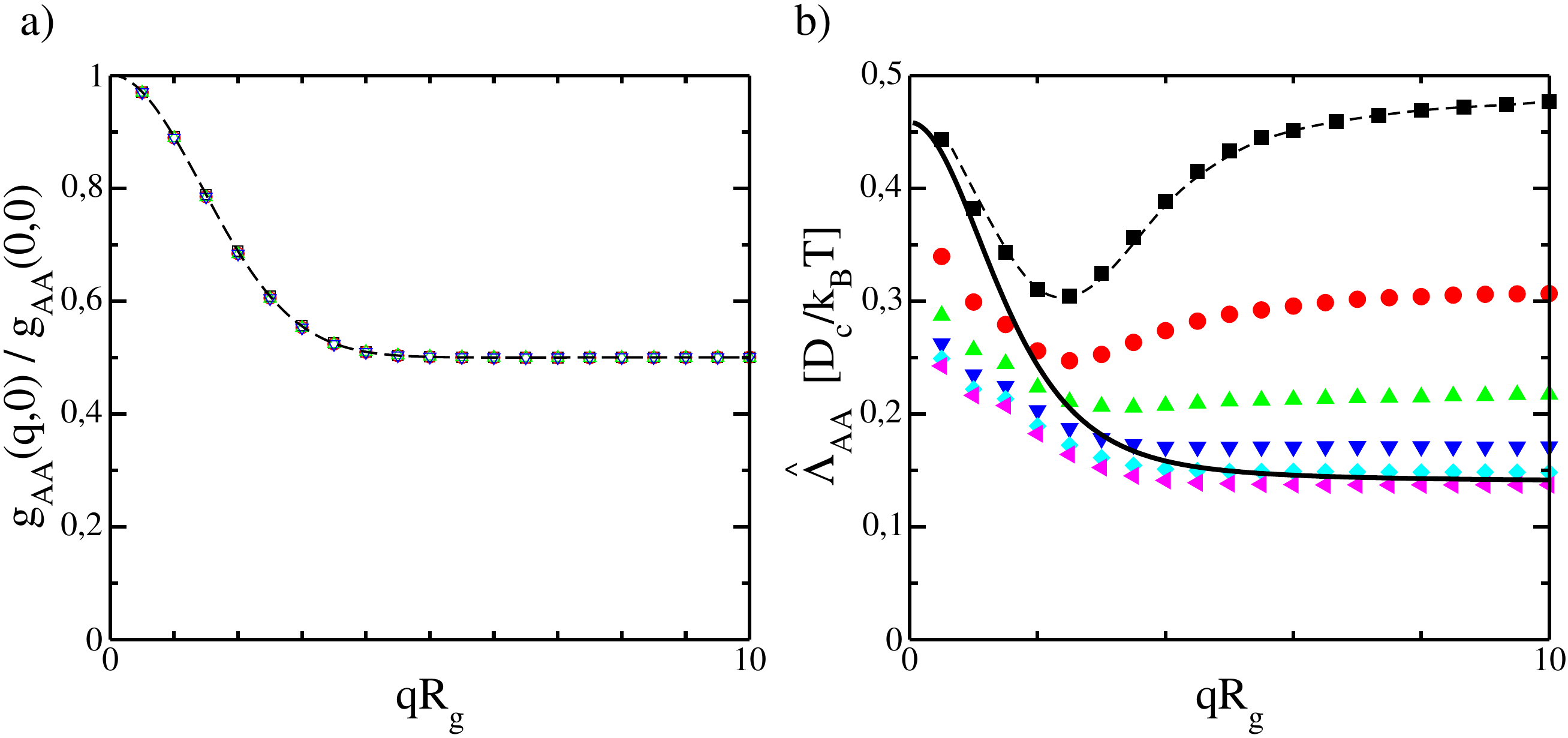}
    \caption{Same as Fig.\ \ref{fig:fig_fullchain}, but now 
    for the first half block (block A) of the chain. 
    a) Normalized static structure factors $g(q,0)$, comparison
    of results from traditional dynamics (dashed line) and 
    modified dynamics with different sets of inverse friction constants
    (symbols).
    b) Corresponding rescaled mobility function, comparison of
    results from traditional dynamics (dashed line), theory for
    Rouse polymers (thick solid line), and modified dynamics with
    inverse relative monomer friction, from top to bottom: 
    $\gamma_m = N\gamma_t$, $N\gamma_t/2$, $N\gamma_t/4$, 
    $N\gamma_t/8$, $N\gamma_t/16$, $N\gamma_t/32$.  }
    \label{fig:fig_a1}
\end{figure} 

The results are shown in Fig.\ \ref{fig:fig_a1}.  Note that
$\hmatL_{AA}(q)=\hmatL_{BB}(q)$ due to symmetry and we also have
$\hmatL_{AB}(q)=\hmatL_{BA}(q)$ and
$\hmatL(q)=\sum_{\alpha\beta}\hmatL_{\alpha\beta}(q)$. Since
$\hmatL(q)$ is known from Fig.\ \ref{fig:fig_fullchain}, it
suffices to plot the data for $\hmatL_{AA}(q)$ here. The same holds
for the static structure factor $g_{\alpha \beta}(q)$.  In Fig.\
\ref{fig:fig_a1}a), we verify that the latter is not affected by
the modified dynamics as expected. The data for the block mobility
functions are given in Fig.\ \ref{fig:fig_a1}b).  At large $q\Rg$,
if one decreases $\gamma_m$, the block mobility function decreases,
and the target value (the value for Rouse polymers) can be matched for
$\gamma_m=N\gamma_t/16$. Different from the total mobility function
$\hmatL$, however, the block mobility function $\hmatL_{AA}$ is also
affected by $\gamma_m$. In regular dynamics ($\gamma_m = \gamma_t N$),
the behavior at small $q \Rg \to 0$ roughly matches that of short
chains. However, if one reduces $\gamma_m$, it becomes smaller and
deviates from the target. Hence it is not possible to match the
kinetics of chain blocks on both short and long length scales in a CG
model with such ultrashort chains, if one uses modified dynamics with
two friction constants.

To analyze this in more detail, we inspect the structure of the block
mobility functions. In Ref.\ \cite{Mantha2020}, we have derived
the following general expressions for $\hmatL_{\alpha \beta}(q)$:
\begin{equation}
\label{eq:lambda_AA}
\hmatL_{AA}(q) = \frac{1}{4k_{B}Tq^{2}N}\left(\frac{g(q,0)}{\tau_R}+\frac{\Delta(q,0)}{\tau_{\Delta}}\right)
\end{equation}
\begin{equation}
\label{eq:lambda_AB}
\hmatL_{AB}(q) = \frac{1}{4k_{B}Tq^{2}N}\left(\frac{g(q,0)}{\tau_R}-\frac{\Delta(q,0)}{\tau_{\Delta}}\right)
\end{equation}
where $\tau_R=\frac{1}{g(q,0)}\int_0^{\infty}dt g(q,t)$,
$\Delta(q,t)=g_{AA}(q,t)+g_{BB}(q,t)-g_{AB}(q,t)-g_{BA}(q,t)$, and
$\tau_{\Delta}=\frac{1}{\Delta(q,0)}\int_0^{\infty}dt \Delta(q,t)$.
Since $g(q,0)$ and $\Delta(q,0)$ are not affected by $\gamma_m$ (shown
in Fig.~\ref{fig:fig_fullchain}a) and Fig.~\ref{fig:fig_a1}a)), the
dependence of  $\tau_R$ and $\tau_{\Delta}$ on $\gamma_m$ will
determine the behavior of $\Lambda_{AA}$ and $\Lambda_{AB}$.
The time scale $\tau_R$ characterizes the dynamics of the whole chain, 
and $\Delta$ characterizes the relaxation dynamics of blocks with respect 
to each other. 

\begin{figure}[tbp]
    \centering
    \includegraphics[width=0.95\textwidth]{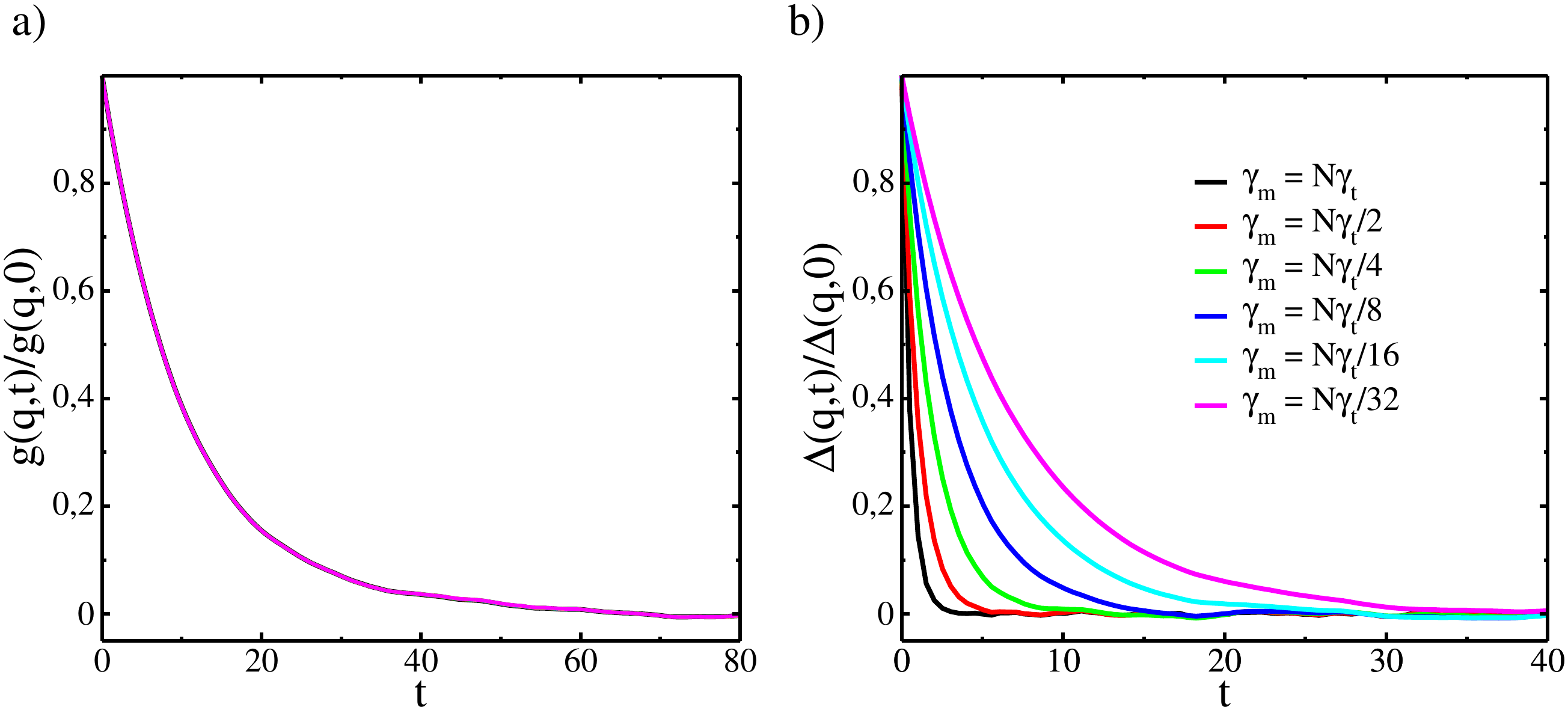}
    \caption{Normalized single chain dynamic structure factor (a) and
    $\Delta(q,t)$ (b) of polymer chain with length $N=4$ obtained from
    modified dynamics at $qR_g = 0.5$. For both figures the monomer
    friction constants $\gamma_m$ is varied as $N\gamma_t$,
    $N\gamma_t/2$, $N\gamma_t/4$, $N\gamma_t/8$, $N\gamma_t/16$, 
    $N\gamma_t/32$. 
    }
    \label{fig:fig_relax}
\end{figure}

Here we focus on the small $q\Rg$ regime.  Fig.~\ref{fig:fig_relax}
shows the normalized single chain dynamic structure factor $g(q,t)$
(a) and the quantity $\Delta(q,t)$ (b) of the CG chains as obtained
from modified dynamics as a function of the simulation time $t$ at $q
\Rg=0.5$. The inverse monomer friction $\gamma_m$ has practically no
effect on the behavior of the single chain dynamic structure factor,
hence the relaxation time $\tau_R$ does not change. For $\Delta(q,t)$,
however, the relaxation slows down with decreasing $\gamma_m$, which
results in an increase of $\tau_{\Delta}$. Combined with the
equations above, we conclude that decreasing $\gamma_m$ will lead to
a decrease in $\Lambda_{AA}$ and a increase in $\Lambda_{AB}$. The
individual blocks relax more slowly and the two blocks move more
cooperatively at small $q \Rg$ if the relative monomer friction is
increased.

\begin{figure}[tbp]
    \centering
    \includegraphics[width=0.95\textwidth]{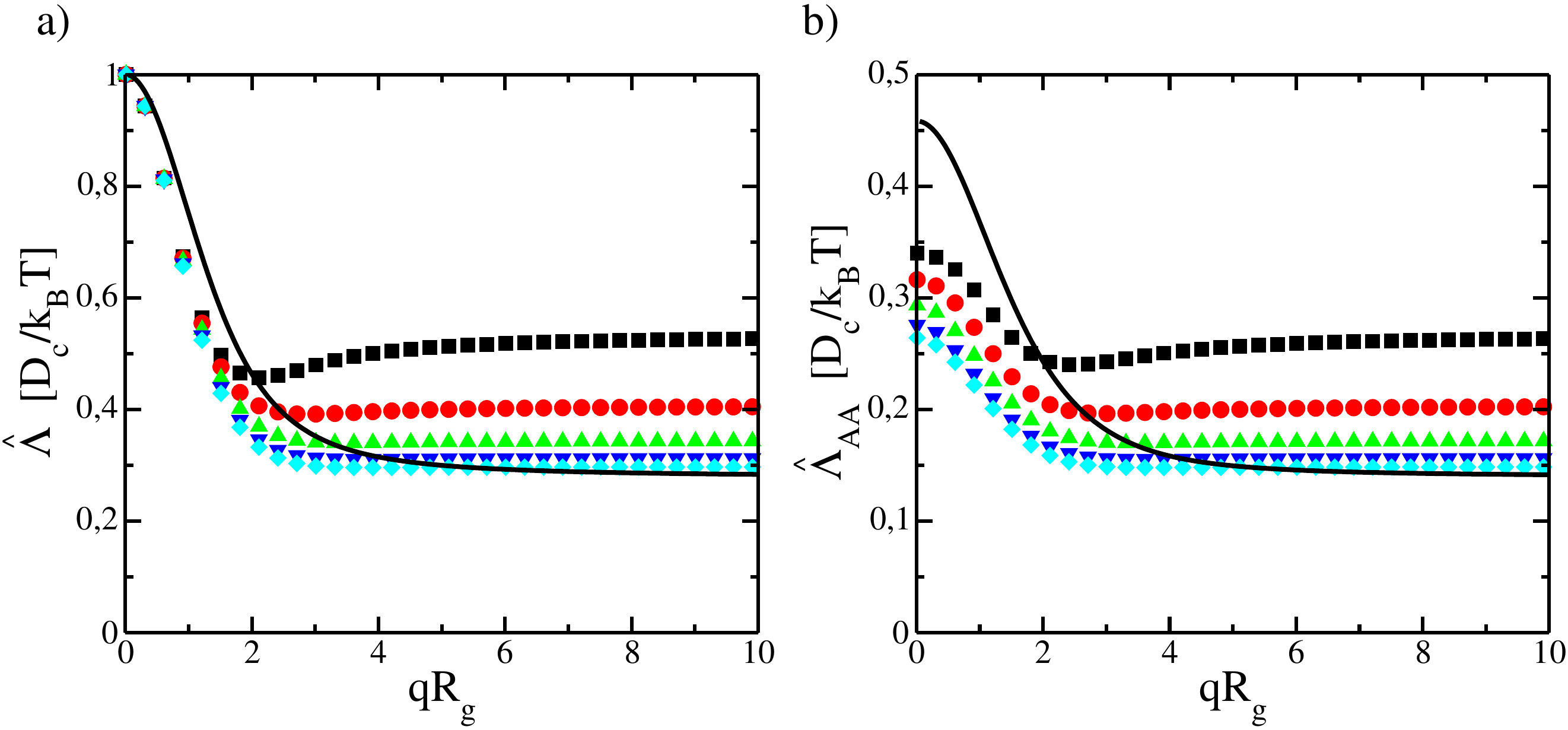}
    \caption{Mobility functions $\Lambda$ (a) and $\hmatL_{AA}$ (b)
    for polymer chains with length $N=4$ from modified dynamics with
    three friction constants. The black solid line shows the target
    function, the mobility function for Rouse polymers. For modified
    dynamics, the inverse total friction constant $\gamma_t$ and the
    inverse relative monomer friction  $\gamma_m=N\gamma_t/16$ are
    kept fixed, and the inverse block friction parameter
    $\gamma_b$ decreases from top to bottom: $N\gamma_t/2$,
    $N\gamma_t/4$, $N\gamma_t/8$, $N\gamma_t/16$, $N\gamma_t/32$.  }
    \label{fig:fig_a2}
\end{figure}

We have tested whether it is possible to decouple the motion of blocks
at small $q \Rg$ by using a more versatile modified dynamics scheme
with three friction constants. To this end, we have adopted the
hierarchical model described in Sec.\ \ref{Sec:Generalization} and
introduced an additional inverse block friction constant $\gamma_b$.
Some representative results are shown in Fig.\ \ref{fig:fig_a2}.  The
black line shows again the target mobility functions. In this example,
we fix the inverse monomer friction parameter at a large value,
$\gamma_m=N\gamma_t/16$, such that relative monomer motions are
largely suppressed, and vary the inverse block friction constant
$\gamma_b$ is varied. As can be seen from Fig.\ \ref{fig:fig_a2},
introducing the hierarchical scheme with three friction constants does
not improve the quality of the mapping. At the level of the block
mobilities, the problems persist, and even the mapping of the total
mobility function (Fig.\ \ref{fig:fig_a2}a)) is not as good as in the
system with two friction constants (Fig.\ \ref{fig:fig_a1}).  We have
explored all possible parameter combinations of $\gamma_b$ and
$\gamma_m$ and did not obtain any better results. Hence we conclude
that dynamic mapping of block copolymers onto tetramers is not
possible, and longer CG chains must be used to model such systems.
Given that the chain length $N=4$ is the minimum chain length for
homopolymer mapping as explained at the beginning of this section,
it is perhaps not surprising that it is too small to map individual
blocks.


\section{Summary and Conclusion}
\label{Sec:Summary}

To summarize, in this paper, we have presented a dynamic
coarse-graining scheme for polymer systems with the goal of mapping the
time scales of local kinetic processes over a large range of relevant
length scales.  The scheme builds on the single-chain mobility matrix,
a wave-vector dependent integrated quantity that is derived from the
single-chain structure factor. We have demonstrated that mobility
functions can be used as sensitive diagnostic tools that highlight the
quality of dynamic mapping schemes for polymers on different length
scales. As an example, we have used them to evaluate extreme
coarse-graining schemes that map long Rouse polymers onto CG
chains with very few effective monomers, and shown that simple time
scale matching fails for large wavevectors $q$. The reason is that in
short chains, the motion of different monomers decouples for large $q
\Rg$, whereas the dynamics remains cooperative in Rouse polymers. As a
remedy, we have proposed a class of modified CG dynamics schemes where the
relative motion of monomers is artificially slowed down, and shown
that this can greatly improve the quality of dynamic mapping of
homopolymers, even if the length of the CG chains is as short as
$N=4$.

We have also investigated the limitations of the method. For
homopolymers, we have established by analytical considerations that
$N=4$ is the minimum CG chain length where consistent dynamic mapping
is possible.  In the case of block copolymers, this still seems too
short and dynamic mapping of symmetric diblock copolymers onto
tetramers was not possible. We found that slowing down the monomers
increases the dynamic correlation between the different blocks in an
undesired way, and it was not possible to find dynamical parameters
that reproduce the mobility matrix function $\hmatL_{\alpha \beta}(q)$
in a satisfactory manner over the whole range of $q$ vectors for CG
chains with $N=4$. We conclude that dynamically consistent ''extreme''
coarse-graining of block copolymers onto CG requires either further
modifications or less extreme coarse-graining (i.e., larger $N$). 
When respecting these limitations, we believe that our scheme can have
a wide range of interesting applications. We have tested it on linear
Rouse polymers, but it can also be applied to polymers in other
dynamic regimes, e.g., entangled polymers, and  to other polymer
architectures. 

Our dynamic coarse-graining scheme is motivated by a Markovian approximation 
to the dynamics (Eq.\ (\ref{eq:DDFT})) that does 
not explicitly account for memory effects in polymer dynamics. Mapping
strategies that target the full frequency dependent mobility
matrix of the GLE, e.g., Eq.\ (\ref{eq:Ks}), should be even more
accurate. However, it will likely not be possible to implement them without
introducing frequency dependent mobility coefficients at the level of 
the CG model as well\cite{Lee2019}, which would greatly reduce the efficiency 
of CG simulations. On the other hand, CG simulations based on modified
dynamics, e.g., Eqs.\ (\ref{eq:dr_modified}) or (\ref{eq:drr_modified}), 
are not much more expensive than regular CG simulations, as they neither
require additional force evaluations, nor extra efforts (storage of 
data, auxiliary variables) to account for memory kernels\cite{li2017computing}.
\Rev{The approach can additionally be motivated by the observation that polymer
DDFTs  based on the Markovian approximation -- when using wave-vector dependent (i.e., nonlocal) mobility functions as in Eq.\ (\ref{eq:DDFT}) -- were found to reproduce kinetic processes in inhomogeneous polymer systems fairly accurately
on time scales well below the Rouse time\cite{Mantha2020}. We have studied this
for chains in the Rouse regime, corresponding investigations of other dynamical regimes are currently under way. 
}

A large number of different internal friction constants can be introduced
following the methods introduced in Section \ref{Sec:Methods} and adjusted 
in order to optimally match the target mobility function. In the present work, 
we have mapped the parameters by straightforward trial and error. In the 
future, it will be desirable to develop more sophisticated iterative mapping
schemes\cite{jung2017iterative,jung2018generalized,Meyer_2019} and/or 
apply machine learning tools\cite{wang2020data} to
optimize the mapping. We believe that such developments will enable
for dynamically accurate large scale simulations of kinetic processes
in inhomogeneous polymer systems by use of extremely coarse-grained
polymer models.

\ack

We thank Shuanhu Qi for valuable discussions.
This work was done within the Collaborative Research Center SFB TRR
146; corresponding financial support was granted by the Deutsche
Forschungsgemeinschaft (DFG) via Grant 233530050.  


\section*{References}

\bibliographystyle{iopart-num}
\bibliography{modified}


\end{document}